\documentclass{article}
\usepackage[utf8]{inputenc}
\usepackage{amsmath}
\usepackage{amssymb}
\usepackage{stmaryrd} 
\usepackage{bm}
\usepackage{cancel}
\usepackage[linesnumbered,ruled,vlined]{algorithm2e}
\usepackage{graphicx}
\usepackage{caption}
\usepackage{subcaption}
\usepackage{multirow}
\usepackage{rotating} 
\usepackage[normalem]{ulem}
\usepackage{mathtools} 
\usepackage{authblk} 

\usepackage{tikz}
\usetikzlibrary{positioning}
\usetikzlibrary{shapes.multipart}
\usetikzlibrary{shapes.geometric}
\usetikzlibrary{arrows.meta}
\usetikzlibrary{backgrounds}
\usetikzlibrary{shapes}
\usetikzlibrary{arrows,decorations.pathmorphing}
\usetikzlibrary{positioning}
\usetikzlibrary{calc}
\usetikzlibrary{intersections}

\newcommand{\comment}[1]{}

\usepackage[sorting=none, backend=bibtex]{biblatex}
\usepackage[colorlinks=true,
	linkcolor=black,
    filecolor=magenta,
    citecolor=magenta,      
    urlcolor=cyan]{hyperref}
\makeatletter
\renewcommand*{\eqref}[1]{%
  \hyperref[{#1}]{\textup{\tagform@{\ref*{#1}}}}%
}
\usepackage{cleveref}
\makeatother
\usepackage{notoccite}
\title{$p$-adaptive algorithms in Discontinuous Galerkin solutions to the time-domain Maxwell's equations}
\author[,1]{Apurva Tiwari\thanks{Corresponding author. \\ \textit{Email addresses: }\href{mailto:apurva.t@aero.iitb.ac.in}{apurva.t@aero.iitb.ac.in}  (A. Tiwari), avijit@aero.iitb.ac.in (A. Chatterjee) }}
\author[1]{Avijit Chatterjee}
\affil[1]{\small \textit{Department of Aerospace Engineering, Indian Institute of Technology Bombay, India}}
\date{}

\begin{document}
\maketitle

\begin{abstract}
The Discontinuous Galerkin time-domain method is well suited for adaptive algorithms to solve the time-domain Maxwell's equations and depends on robust and economically computable drivers. Adaptive algorithms utilize local indicators to dynamically identify regions and assign spatial operators of varying accuracy in the computational domain. This work identifies requisite properties of adaptivity drivers and develops two methods, a feature-based method guided by gradients of local field, and another  utilizing the divergence error often found in numerical solution to the time-domain Maxwell's equations. Results for canonical testcases of electromagnetic scattering are presented, highlighting key characteristics of both methods, and their computational performance.
\end{abstract}

\section{Introduction}
Mesh adaptive methods aim to efficiently allocate discrete degrees of freedom throughout a computational space. The finite element method and development of the popular Discontinuous Galerkin Finite Element Method (DGFEM) provides flexibility with non-conforming meshes having hanging nodes, and facilitates $p$-anisotropy, \textit{i.e.} variation of degree of the basis polynomials across cells. Adaptive methods have been widely studied with initial ideas appearing in the works of Babu\v{s}ka and co-workers \cite{Babuska,Babuska1981,Babuska1986}. Broadly, adaptive methods are classified into three categories: those based on residuals or local error estimation \cite{Kompenhans2016a, KompenhansRubio, Syrakos2012}, adjoint-based methods that estimate the error in a particular simulation output of interest as an objective functional \cite{Wang2009, Pierce2004, Venditti2002, Hartmann2002} and, the feature-based methods, recognized by easily computable indicators, often encoding identifiable physical quantities to be accurately resolved \cite{Aftosmis1994,Remacle2003,Hernandez1997,Luo2003}. 
Feature-based methods use readily computable response variables and their gradients like density or pressure gradients are classically approaches to identify regions of interest. The primary limitation these techniues is that the identified features may not show a direct relation with the underlying numerical discretization errors \cite{Kompenhans2016a}. Also, the subjectivity involved due to their heuristic nature requires an expert user to devise effective feature-based schemes \cite{Naddei}.
Adjoint-based methods eliminate these drawbacks to an extent, relying on numerical approximations of the adjoint problem to compute the error in a target functional. Despite being theoretically sound, these methods have not found widespread industrial application, owing to their high computational cost \cite{Fraysse2012a,Rueda-Ramirez2019}. Residual or error-based methods work towards making the mesh richer by estimating local sources of error. This approach has been shown to be computationally more economical than adjoint methods, and are autonomous \cite{KompenhansRubio}. Error estimation driven adaptive methods have found a lot of interest and a wide range of error estimation techniques and applications can be found in literature \cite{Kompenhans2016a,Gratsch2005,AinsworthMono}. In hyperbolic problems, the two relevant numerical errors are discretization and truncation errors. The discretization error is advected downstream from under-resolved regions in space, unlike the truncation error, that acts as a local source \cite{Rueda-Ramirez2019}. The truncation error is related to the discretization error through the Discretization Error Transport equation (DETE) and forms the case for truncation error driven adaptive methods in hyperbolic problems \cite{Roy}. Truncation error estimation, or $\tau$-estimation based adaptive methods have been successfully used in case of wave-dominated problems \cite{Kompenhans2016a,Syrakos2012}. Hence, the effectiveness of adaptive methods depends largely on developing robust error indicators and often, the utility of these methods is limited by the following challenges, as is highlighted in \cite{Zienkiewicz1987}:
\begin{enumerate}
\item Prohibitive cost of computing effective error estimators.
\item The practical difficulty of incorporating a fully adaptive routine in an existing code structure.
\end{enumerate}
\par Given that the conventional numerical framework used in solid mechanics for long is the finite element method, we investigate advancements in the development of well-established refinement indicators in problems of linear elasticity to identify requisite properties of such indicators for possible use in time-domain electromagnetics. One of the most accepted physical quantities that drivers are based on, is the strain energy or strain energy density (SED). Strain energy being used in various forms in adaptive methods is standard and has a long history. One of the first uses of strain energy as an error sensor was shown by Melosh and Marcal \cite{Melosh1977} and was further developed in later works by Shephard \cite{Shephard1980} and Botkin \cite{Botkin1985}, where variation in SED was used to estimate local error. Hern\'{a}ndez \cite{Hernandez1997} used error in SED to drive the mesh refinement, and Luo \cite{Luo2003} used its gradient. The use of energy to drive an adaptive method is also seen in other disciplines, for example, Heid et.\ al.\ in \cite{Heid2021} used energy based adaptivity to model the steady states of Boson Einstein condensates consisting of a collection of bosonic quantum particles. The working principle is that for a structure with a given load path, at the equilibrium state, the total potential energy is definite. This has been presented in the context of the minimization of total potential energy principle in structural mechanics in \cite{Luo2003}. Therein, the theoretical basis to use the SED is that the numerical strain energy must converge to its actual value with finer meshes, and gradient of SED is used as a heuristic, feature-based driver. This is distinguished from schemes using error in SED instead.
Hern\'{a}ndez et. al. in \cite{Hernandez1997}, showed that one of the most widely used measure of error in linear elasticity, \textit{i.e.} the energy norm, from the SED. This led to the use of error in SED as a proxy to discretization error, and a subsequent error-driven adaptation scheme \cite{Hernandez1997}. A strain energy-based refinement indicator also possesses characteristics like being a scalar and co-ordinate independent, which are of practical significance. Thus, a versatile and robust indicator for adaptive methods is found in strain energy. 
\par  In this paper, we seek a successor of the strain energy based indicators established in solid mechanics, to compute $p$-adaptive DGTD solutions to the time-domain Maxwell's equations. We investigate established strain energy based indicators to identify requisite properties of a prospective indicator, and use these as guidelines to develop a driver for $p$-adaptive methods in CEM. Specifically, we are interested in wave scattering problems modelled using the time-domain Maxwell's equations. Electromagnetic (EM) energy in CEM has been used in literature for various purposes.  Henrotte \cite{Henrotte2007} presented an energy-based representation of the Maxwell system to develop weak formulations in CEM which enabled coupling multi-physics sub-problems to establish a global energy balance. Further applications, namely in model order reduction techniques, and parameter identification methods have also been shown \cite{Henrotte2007, Antoulas2000}.
\par On the subject of adaptive methods, there has been recent interest in utilizing EM energy as a driver. Feature-based heuristic adaptive algorithms have been proposed for the time-domain Maxwell's equations, set in the Discontinuous Galerkin time domain (DGTD) framework, driven by gradient of EM energy \cite{Yan2017b}, and expansion co-efficients of modal basis polynomials used to represent the solution \cite{Yan2017c}.  $hp$-refinement techniques have also been developed based on energy-norm of the discretization error \cite{Garcia-Castillo2008}, and extended to goal-oriented adaptivity based on the minimization of error in scattering parameters in waveguides \cite{Garcia-Castillo2008,Pardo2007}. However, as pointed out in \cite{Burg2013}, these methods introduce a global optimization scheme, which proves to be computationally expensive since it solves a finer version of a given problem. This essentially highlights a key limitation of error estimation: that estimating the error, requires estimating a finer solution, or a reference value since it is usually unknown. The procedure of adopting an enhanced or smoothed solution on the basis of a coarse solution is well established in finite element methods \cite{Hernandez1997,Tessler,Donzelli1992}.
There is abundant literature available for $p$- and $hp$- adaptive error-estimation methods in elliptic problems \cite{Houston,Ainsworth1998,Eibner2007,Heuveline2003,Rachowicz2006}, but those for the hyperbolic time-domain Maxwell's equations are limited. We find a gap in literature studying a non-heuristic adaptive method for the time-domain Maxwell's equations building upon the idea of utilizing EM energy as a driver. 
\par We analyze an energy-based $p$-adaptive algorithm, borrowing key ideas from \cite{Luo2003,Luo2010} originally developed based on SED for problems in structural mechanics for problems in linear elasticity, exploring the extent they can be emulated in CEM to solve the time-domain Maxwell's equations. We further the analysis by showing that for the purpose of adaptivity indicators, EM energy in CEM does not assume the same role as strain energy in solid mechanics. We also show how the pitfalls of EM energy as an indicator can instead be overcome with a re-interpretation of divergence error, often found in numerical solutions of the time-domain Maxwell's equations, in non-finite difference time domain methods. Ref. \cite{ApurvaDiv} introduces the idea of utilizing divergence error as an indicator in adaptive algorithms, combining the ease of computation of feature-based methods and the rigour and autonomy of error estimation based methods. 

\section{Discontinuous Galerkin time domain method}
\label{sec:DGTD}
This section briefly outlines the discontinuous Galerkin time domain method. \par Consider a system of hyperbolic conservation laws,
\begin{equation}
\frac{\partial \mathbf{u}}{\partial t}+\nabla\cdot\mathcal{F} = \mathbf{s}
\label{eq:cons}
\end{equation}
where, $\mathbf{u}=\left[q_1, \cdots, q_n\right]^T$ is a vector of conserved variables,  $\mathcal{F}$ is the flux tensor depending on $q$, and $s$ is a source term.
This system is representative, among others, of the time domain Maxwell's curl equations. For instance, consider the vector components in the 2D transverse magnetic ($TM_z$) mode, given by
\begin{equation}
\mathbf{u}=\left(
\begin{array}{c}
B_x \\ B_y \\ D_z
\end{array}
\right), \hspace*{0.3cm}
\mathbf{f}=\left(
\begin{array}{c}
0 \\ -D_z/\epsilon \\ -B_y/\mu
\end{array}
\right), \hspace*{0.3cm}
\mathbf{g}=\left(
\begin{array}{c}
D_z/\epsilon \\ 0 \\ B_x/\mu
\end{array}
\right), \hspace*{0.3cm}
\mathbf{s}=\left(
\begin{array}{c}
0 \\ 0 \\ -J_{iz}
\end{array}
\right).
\label{eq:TM}
\end{equation} 
and in the 2D transverse electric ($TE_z$) mode given by,
\begin{equation}
\mathbf{u}=\left(
\begin{array}{c}
B_z \\ D_x \\ D_y
\end{array}
\right), \hspace*{0.25cm}
\mathbf{f}=\left(
\begin{array}{c}
D_z/\epsilon \\ 0 \\ B_z/\mu
\end{array}
\right), \hspace*{0.25cm}
\mathbf{g}=\left(
\begin{array}{c}
-D_x/\epsilon \\ -B_z/\mu \\ 0 
\end{array}
\right), \hspace*{0.25cm}
\mathbf{s}=\left(
\begin{array}{c}
0 \\ -J_{ix} \\ -J_{iy}
\end{array}
\right).
\end{equation}
Here, $\mathcal{F}=\left[\mathbf{f},\mathbf{g}\right]$ consists corresponding flux vectors in the $x$ and $y$ directions. To close the system, constituent relations $\mathbf{B}=\mu\mathbf{H}, \mathbf{D}=\epsilon \mathbf{E}$ are used.
The spatial domain $\Omega$ is triangulated as $K$ elements, $\Omega\simeq\Omega_h=\bigcup_{k=1}^K\mathsf{D}^k$, and  boundary $\partial\Omega_h$. Element $\mathsf{D}^k$ is a straight-sided triangle with the triangulation taken to be geometrically conforming.
\par The solution $\mathbf{u}$ is approximated locally as a polynomial expansion using $p$-th degree nodal basis functions, defined on element $\mathsf{D}^k$. This variational method involves multiplying the conservation law by a test function and integrating by parts. Choosing the test function to be the local basis functions leads to the Galerkin formulation,
\begin{equation}
\int_{\mathsf{D}^k}\left(\frac{\partial \mathbf{u}_{h,p}^k}{\partial t}+
\nabla\cdot\mathcal{F}_{h,p}^k\right)\ell_i^k(\bm{x})d\bm{x}=\int_{\partial\mathsf{D}^k}\bm{\hat{n}}\cdot\left[\mathcal{F}_{h,p}^k-\mathcal{F}^*\right]\ell_i^k(\bm{x})d\bm{x},
\label{eq:strongform}
\end{equation}
A $p$-th order multidimensional Lagrange polynomial $\ell_i^k$ is constructed with nodes $x_i$ on element $\mathsf{D}^k$. Since the discontinuous Galerkin schemes allows for discontinuities at element interfaces, this leads to multi-valued functions at the the edges. An upwind flux $\mathcal{F^*}$ is used to resolve the multi-valuedness. $\bm{\hat{n}}$ is the local outward pointing normal to the element edge $\partial \mathsf{D}^k$. For a thorough description of the details, we refer the reader to \cite{Hesthaven}.
\par
\begin{figure}
\centering
\scalebox{0.75}{
\begin{tikzpicture}
\tikzstyle{obj}  = [circle, minimum width=6pt, fill, inner sep=0pt]

\node[isosceles triangle,
    draw,
    fill=cyan!10,
    rotate=-30,
    minimum size =4cm] (Tl) at (0,0){};
    
\node[isosceles triangle,
    draw,
    fill=blue!10,
    rotate=150,
    minimum size =4cm] (Tr) at (4.5,1.5) {};

\node[obj] (tl1) [above=0cm of Tl.apex, anchor=center] {};    
\node[obj] (tl2) [above=0cm of Tl.20, anchor=center] {};
\node[obj] (tl3) [above=0cm of Tl.80, anchor=center] {};
\node[obj] (tl4) [above=0cm of Tl.left corner, anchor=center] {};
\node[obj] (tl5) [above=0cm of Tl.150, anchor=center] {};
\node[obj] (tl6) [above=0cm of Tl.210, anchor=center] {};
\node[] (p3) [above left= of tl6, anchor=center] {\huge $\bm{p=3}$};
\node[obj] (tl7) [above=0cm of Tl.right corner, anchor=center] {};
\node[obj] (tl8) [above=0cm of Tl.280, anchor=center] {};
\node[obj] (tl9) [above=0cm of Tl.340, anchor=center] {};

\path[name path=line 1] (Tl.left side) -- (tl7);
\path[name path=line 2] (tl5) -- (tl9);
\path [name intersections={of=line 1 and line 2,by=tl10}];
\node[obj, opacity=0.5] at (tl10) {};

\node[obj] (tr1) [above=0cm of Tr.apex, anchor=center] {};
\node[obj] (tr2) [above=0cm of Tr.10, anchor=center] {};
\node[obj] (tr3) [above=0cm of Tr.left side, anchor=center] {};
\node[obj] (tr4) [above=0cm of Tr.100, anchor=center] {};
\node[obj] (tr5) [above=0cm of Tr.left corner, anchor=center] {};
\node[obj] (tr6) [above=0cm of Tr.140, anchor=center] {};
\node[obj] (tr7) [above=0cm of Tr.lower side, anchor=center] {};
\node[] (p4) [right=2cm of tr7, anchor=center] {\huge $\bm{p=4}$};
\node[obj] (tr8) [above=0cm of Tr.220, anchor=center] {};
\node[obj] (tr9) [above=0cm of Tr.right corner, anchor=center] {};
\node[obj] (tr10) [above=0cm of Tr.260, anchor=center] {};
\node[obj] (tr11) [above=0cm of Tr.right side, anchor=center] {};
\node[obj] (tr12) [above=0cm of Tr.350, anchor=center] {};
\path[name path=line 1] (tr3) -- (tr11);
\path[name path=line 2] (Tr.apex) -- (Tr.lower side);
\path [name intersections={of=line 1 and line 2,by=tr13}];
\path[name path=line 1] (tr3) -- (tr7);
\path[name path=line 2] (tr5) -- (Tr.right side);
\path [name intersections={of=line 1 and line 2,by=tr14}];
\node[obj, opacity=0.5] at (tr13) {};
\node[obj, opacity=0.5] at (tr14) {};

\path[name path=line 1] (tr7) -- (tr11);
\path[name path=line 2] (Tr.left side) -- (Tr.right corner);
\path [name intersections={of=line 1 and line 2,by=tr15}];
\node[obj, opacity=0.5] at (tr15) {};

\draw[->, opacity=0.7] (tl3) -- ++(30:1.25cm) node[right] {\huge ?};
\draw[->, opacity=0.7] (tl2) -- ++(30:1.25cm) node[right] {\huge ?};
\draw[->, opacity=0.7] (tr2) -- ++(-150:1.25cm) node[left] {\huge ?};
\draw[->, opacity=0.7] (tr3) -- ++(-150:1.25cm) node[left] {\huge ?};
\draw[->, opacity=0.7] (tr4) -- ++(-150:1.25cm) node[left] {\huge ?};
\end{tikzpicture}}
\caption{``Misaligned" neighbouring nodes at common edges in a $p$-anisotropic nodal DG framework}
\label{fig:misaligned_neighbours}
\end{figure}
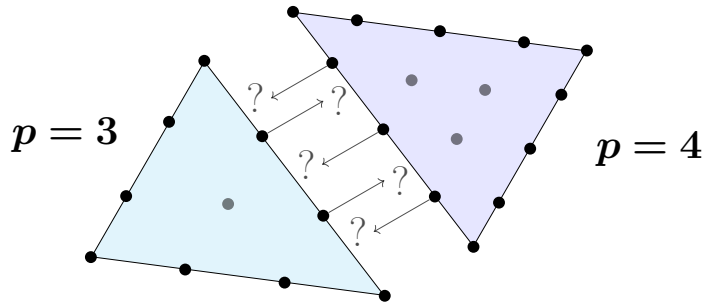

 In a nodal DG framework, nodes are spaced according to the degree of the quadrature employed, which varies across elements due to the $p$-anisotropy. This gives rise to situations like the one shown in fig. \ref{fig:misaligned_neighbours}. Neighbouring cells with dissimilar $p$, do not have the solution available at the same physical positions. This requires evaluating the solution by interpolating it using its polynomial expansion \cite{Moxey}. Every update in the $p$-distribution may requrie such computation for each pair of common edges. Another concern is truncation of the outer boundaries of the computational domain $\Omega_h$. For scattering problems, we require that the scattered field dampens as it moves sufficiently far away from the scatterer to safely truncate the domain \cite{Hesthaven}. The computational domain consists of a  surrounding perfectly matched layer (PML) along the outer boundaries such that they do not produce spurious oscillations at their interface with the inner domain \cite{Abarbanel1998}. Testcases in the results presented, mention geometry of the PML layer used in them. A practical overhead of the $p$-anisotropy is the increased bookkeeping of the field vectors, boundaries, element matrices etc. Given the unequal sizes of local solution vectors across elements, indexing into them needs special treatment too. 

\section{Energy driven adaptivity}
\label{sec:Energy}
\subsection{Role of strain energy in solid mechanics}
\label{sec:SEDprops}
In solid mechanics, a widely accepted driver for adaptive methods is the strain energy. Strain energy and its density have been used in various forms to drive adaptive methods. Error in SED as a local error estimate is well established \cite{Hernandez1997, Botkin, Shephard1980, Melosh1977}, and is coupled with the procedure of treating an enhanced field obtained from a coarse discrete solution as reference to estimate error \cite{Donzelli,Tessler}, to devise an adaptive scheme. Hern\'{a}ndez presented this process  \cite{Hernandez1997} and derived a relation between the energy norm of the discretization error in displacements $||\tilde{e}||_E$, and the error in SED as,
\begin{equation}
\label{eq:SEDreln}
||\tilde{e}||_E^2=2\cdot\int_V|\Psi^*-\tilde{\Psi}|dv
\end{equation}
where $\Psi^*$ is taken to be the true SED $\Psi$, and $\tilde{\Psi}$ is the SED computed from the numerically obtained field. $dv$ is an elemental volume in a finite element discretization. The enhanced field $\Psi^*$ is suggested to be obtained from $\tilde{\Psi}$ using simple averaging at nodes, or various projection methods. 

\par  Gradient of SED is also used to compute a heuristic refinement criterion \cite{Luo2003,Luo2010}. Luo et. al. developed gradient-based methods set in the element-free Galerkin framework, based on the on the principle that higher gradients require finer meshes. A basic requirement of convergence of the numerical SED is applied. A given structure under a set of given loads corresponds to a definite value of the total strain energy. The rationale also includes the fact that SED comprehensively embeds information of other fields like the displacement, stress, material properties etc. 
To summarize, SED forms a basis in both error-based and feature-based adaptive methods in solid mechanics, mainly due to the following factors:
\begin{enumerate}
\label{enum:list}
\item SED is a scalar and independent of the co-ordinate system used.
\item The numerical strain energy converges to a definite value. 
\item Information regarding other fields like the displacement and stress, are embedded into the SED.
\item A measure of error can be derived from the SED, making it an error estimator.
\end{enumerate}
 The first 3 points are common to both feature-based and error-based drivers. The last point admitting SED as an error estimator enables a ``blackbox" operation \cite{Botkin}.
 
\subsection{EM energy as driver}
In contrast to solid mechanics, adaptive methods in advection dominated problems in CEM, require tracking oscillatory waves. Finer meshes are needed to resolve drastic spatial variation in such fields \cite{Yan2017b,Yan2017}. The variation in the fields is measured by the gradient of the local EM power density $\nabla\left(\epsilon||\mathbf{E}||^2+\mu||\mathbf{H}||^2\right)$, or that of the local electric field $E_z$, in the $TM_z$ case, as suggested in \cite{Yan2017b}. This results in a feature-based scheme driven by EM energy. 
\par We analyze EM energy as a driver, considering  the properties of SED listed in sec. \ref{sec:SEDprops} as guidelines. Like SED, EM energy is obviously a scalar and independent of co-ordinate systems. We now show that EM energy indeed converges to a definite value with finer meshes. 

\subsection{Asymptotic convergence of EM energy}
\label{sec:Convergence_EME}
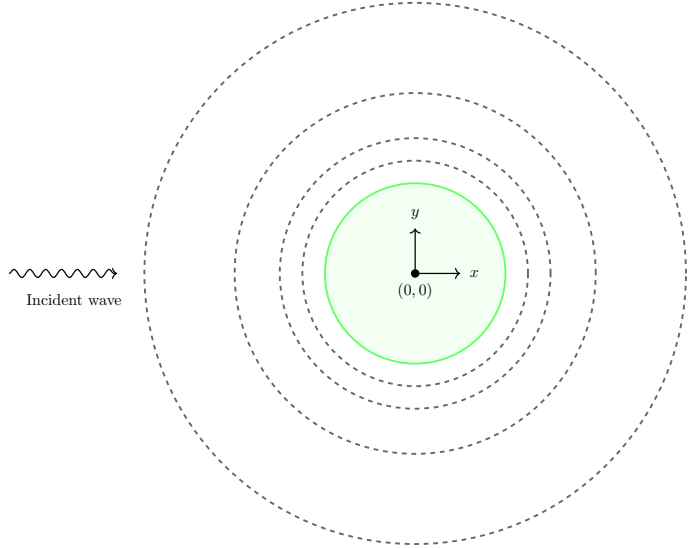
\begin{figure}
\centering
\scalebox{0.6}{
\begin{tikzpicture}[
photon/.style={decorate, decoration={snake}, draw=black},
squaregreennode/.style={circle, draw=black, fill=black, very thick, scale=0.5},
roundnode/.style={circle, draw=green!60, fill=green!5, very thick, minimum size=40mm},
roundnode2/.style={draw,shape=circle,fill=black,scale=0.5},
dashnode125/.style={dashed,circle, draw=black!60,very thick, minimum size=50mm},
dashnode15/.style={dashed,circle, draw=black!60,very thick, minimum size=60mm},
dashnode2/.style={dashed,circle, draw=black!60,very thick, minimum size=80mm},
dashnode3/.style={dashed,circle, draw=black!60,very thick, minimum size=120mm},
coord/.style={draw,shape=circle,fill=black,scale=0}
]

\node[roundnode]      (scat)         {};
\node[roundnode2,
label={[label distance=1.0cm]0:$x$},
label={[label distance=1.0cm]90:$y$},
label={[label distance=0.0cm]270:$\left(0,0\right)$},
label={[label distance=6.3cm]183:Incident wave},draw]      (orig) {};
\node[dashnode125]        (r125)        {};
\node[dashnode15]        (r15)        {};
\node[dashnode2]        (r2)        {};
\node[dashnode3]     (r2)        {};

\draw[black, thick,->] (orig) -- (1,0);
\draw[thick,->,photon] (-9,0) -- (-6.6,0) ;
\draw[black, thick,->] (orig) -- (0,1);
\end{tikzpicture}}\\
\caption{Enclosing surfaces (dashed) around scattering cylinder (solid)}
\label{fig:enclose}
\end{figure}

We use the Richardson extrapolation and grid convergence criteria to quantify the asymptotic behaviour of EM energy. Consider a function $f$, computed using two meshes, a fine and a coarse, their sizes denoted by $h_1$, $h_2$ and the computed values by $f_1$, $f_2$. This makes the grid refinement ratio $s=h_2/h_1$. The continuum value $f_{h=0}$ can be approximated using a generalization of Richardson extrapolation \cite{NASA_GridConv} for a $p$-th order method as,
\begin{equation}
f_{h=0} \cong  f_1 + \frac{f_1-f_2}{s^p-1}.
\end{equation}
Roache \cite{Roache_uncertainty} suggested a grid convergence index for uniform reporting of grid convergence studies. Assuming $3$ meshes denoted by subscripts $1, 2, 3$; $1$ being the finest, the grid convergence index between grids $1$ and $2$ is given by
\begin{equation}
GCI_{12} = \frac{F_s|\epsilon_{12}|}{s^p-1},
\label{eq:def_GCI}
\end{equation}
where $\epsilon_{12}$ is the relative error $\left(f_1-f_2\right)/f_1$. $F_s$ is a factor of safety usually taken to be $1.25$ when comparing three or more grids. This index shows how far the solution is from its asymptotic value. In the asymptotic range, consecutive $GCI$s follow the relation $GCI_{23}=s^pGCI_{12}$, or in other words, the ratio,
\begin{equation}
\mathcal{R}=\frac{GCI_{23}}{s^pGCI_{12}} = 1.
\label{eq:def_R}
\end{equation}
Following figures \ref{fig:r_conv_integral}-\ref{fig:GCI_aR} establish asymptotic convergence of EM energy density in a scattering problem. Fig. \ref{fig:enclose} shows a schematic with a circular cylinder of radius $r=1$ as the scatterer, with a monochromatic $TM_z$ incident plane wave travelling along the $+x$-axis.  The scatterer is enclosed by $4$ other concentric circles of radii $r=1.25,1.5,2,3$ to mark locations where the energy is recorded. Fig.\ \ref{fig:r_conv_integral} shows the asymptotic behaviour of energy density, approximated with the scattered electric field as integral of $E_z^2$ along the  scatterer's surface and the enclosing circles of various radii. Since $E_z$ is harmonic in time, each plot in fig.\ \ref{fig:r_conv_integral} shows values $\max_t E_z^2(r,\theta,t)$ computed for various $p=[2,\cdots,5]$ with refined meshes denoted by the no. of cells $K_{cyl}$ the scattering surface is divided into. For reference, the Richardson extrapolated value for the data corresponding to $p=5$ is shown with a solid black line. Energy of the signals corresponding to all $p=[2,\cdots,5]$, and at all locations recorded, shows convergence to the extrapolated value asymptotically. The local behaviour is shown in fig.\ \ref{fig:rconv_local}, where the scattered field is plotted against viewing angle $\theta$, measured anticlockwise from $+x$-axis; for different levels of mesh refinement. Each row corresponds to a particular $p$ and each column, an enclosing circle. Each subplot shows data corresponding to various $K_{cyl}$ levels of mesh refinement. Beginning with an under-resolved mesh with $K_{cyl}=16$, plots with finer meshes are overlaid. The mutual agreement between plots corresponding to all mesh levels (except the under-resolved one) shows convergence of the local field $E_z$ and of the local EM  energy in turn. This is quantified in fig.\ \ref{fig:GCI_aR} that uses the three finest meshes $K_{cyl}=[32,64,128]$ to plot GCI and the ratio $\mathcal{R}$ between successive GCIs as 
defined in eqns.\ \eqref{eq:def_GCI} and \eqref{eq:def_R}. Therefore, we find that the local field and its energy show an asymptotic convergence to a definite value, much like its solid mechanics counterpart, strain energy density.
\par As listed in sec.\ \ref{sec:SEDprops}, SED is a scalar, co-ordinate independent and is clearly headed towards a definite state. With EM energy showing similar traits, we develop an adaptive scheme driven by the gradient of EM energy. An important distinction must be made between the nature of computing the SED and EM energy. The solution obtained in solid mechanics is expressed in terms of nodal displacements, and obtaining strain involves their derivatives. The SED $\tilde{\Psi}$ computed using numerically obtained nodal displacements $\delta$ \cite{Hernandez1997}, is given by,

\begin{equation}
\tilde{\Psi}=\frac{1}{2}\delta^T \mathbf{B}^T\mathbf{D} \mathbf{B}^T\delta,
\label{eq:gradinfo}
\end{equation}
where $\mathbf{D}$ is the elasticity matrix, and $\mathbf{B}$ is the matrix relating strain and nodal displacements $\delta$ via a differential operator matrix relating the strains to the continuous displacement field. Therefore, the SED involves information of the derivatives of the solution whereas in the case of CEM, computing the energy of a field does not add any additional smoothness information. This can be seen from the expression for the total energy stored in electromagnetic fields $U_{em}$ \cite{Griffiths}, given by,

\begin{equation}
U_{em}=\frac{1}{2}\int \left(\epsilon E^2 + \frac{1}{\mu} B^2\right) dv 
\end{equation} 

Moreover, there exists no relation between error in EM energy or its gradient, and the underlying numerical errors, as opposed to SED in solid mechanics. This limits the use of EM energy to only feature-based adaptive schemes. We present such a feature-based adaptive strategy based on the gradient of local EM energy. The choice of gradient of a local field as a driver has been used in CEM too, in reference \cite{Yan2017b}, to determine levels of refinement heuristically, for similar problems as the one presented in this work.

\begin{figure}
\centering
i\includegraphics[width=\textwidth]{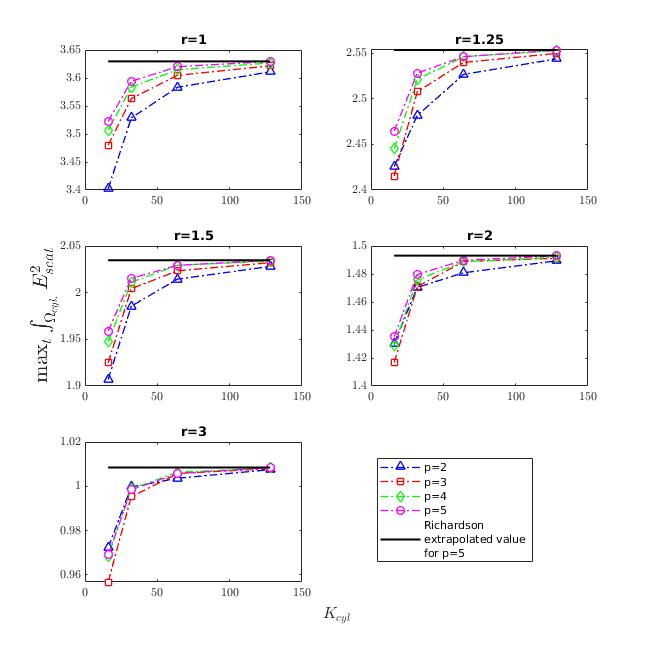}
\caption{Asymptotic convergence of time-averaged $\int_{\Omega_{cyl}}\left(E_z^s\right)^2$ \textit{vs.} no. of elements $K_{cyl}$ on the surface of the scattering cylinder.}
\label{fig:r_conv_integral}
\end{figure}

\begin{sidewaysfigure}
\centering
\includegraphics[width=1.2\textwidth]{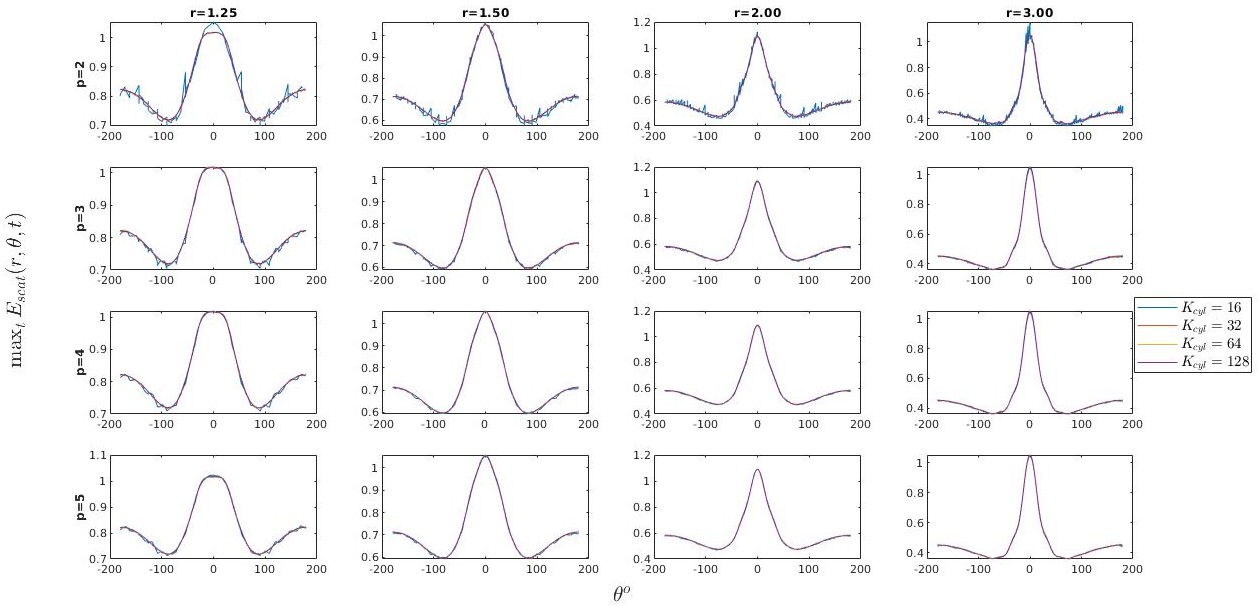}
\caption{Asymptotic convergence of time-averaged, local $\max_t E_z^s$ \textit{vs.} viewing angle $\theta$.}
\label{fig:rconv_local}
\end{sidewaysfigure}

\begin{sidewaysfigure}
\flushleft
\includegraphics[width=1.2\textwidth]{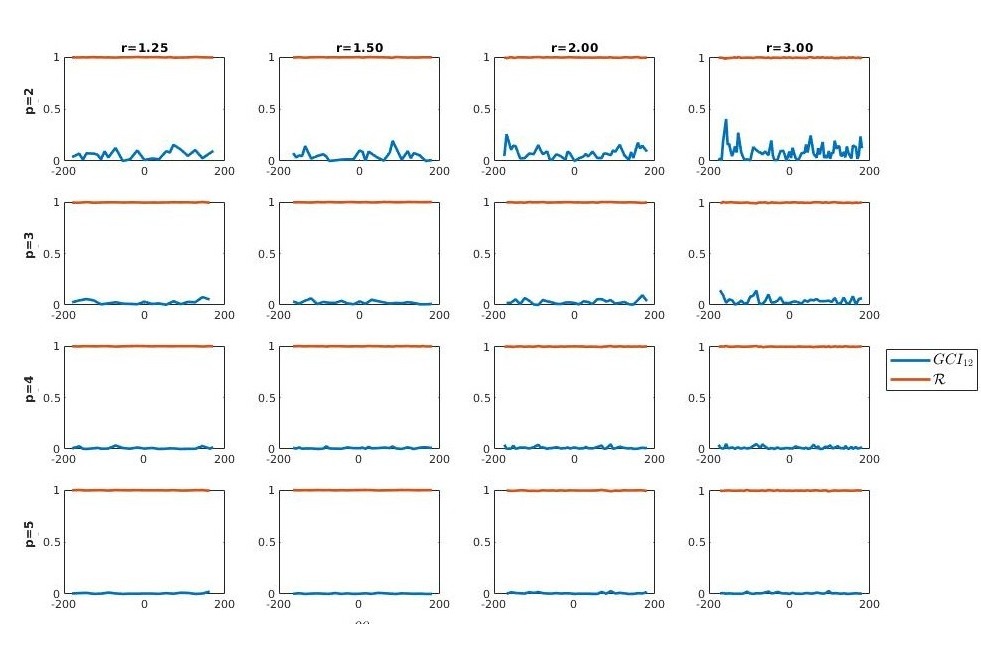}
\caption{GCI (in $\%$) and $\mathcal{R}$ of time-averaged, local $E_z^s$ \textit{vs.} viewing angle $\theta$.}
\label{fig:GCI_aR}
\end{sidewaysfigure}

\section{Divergence error based adaptivity}
\label{sec:Divergence}
 In this section, we summarize a re-interpretation of a known numerical error arising when solving the time-domain Maxwell's equations, the divergence error, as an error indicator presented by us earlier in ref. \cite{ApurvaDiv}. Numerical solutions of the time-domain Maxwell's equations, when obtained with methods other than the finite-difference time-domain (FDTD) methods, consist of a numerical divergence error originating from the treatment of the constraint on divergence posed by the Gauss laws. Due to the act of discretization with spatial operators having finite accuracy, the solenoidal condition on the fields is not met exactly. This results in the evolution of divergence in an initially solenoidal field, which in the continuous system is zero (in absence of sources), giving rise to numerical divergence error.
  There is abundant literature on the subject of divergence cleaning techniques and Gauss' constraint satisfying methods \cite{CockburnLDG,Li2012,Yakovlev2013,Chandrashekar2019,Munz,Assous1993}. 
But these are only used when the underlying physics requires exactly meeting the divergence constraint, for example in magnetohydrodynamics. Apart from this exception, divergence errors accruing in conservative higher-order formulations do not significantly impact the overall accuracy of the solution \cite{Munz,Cioni} and so, are often disregarded in practice.  In \cite{Cioni}, Cioni et. al used a mixed finite volume/finite element method to show that divergence error, despite being linked to the accuracy of the solver and the underlying discretization, does not hamper the formal accuracy of the solution. 
\par Consider a subsystem of eq. \eqref{eq:cons} with the $TM_z$ mode given by eq. \eqref{eq:TM}, consisting of only in-plane components $B_x$ and $B_y$, such that
\begin{equation}
\mathbf{U}=
\begin{pmatrix}
B_x \\ B_y 
\end{pmatrix}
=\mathbf{B}, \hspace*{0.3cm}
\mathbf{F}=
\begin{pmatrix}
0 \\ -D_z/\epsilon 
\end{pmatrix}\hat{i}
, \hspace*{0.3cm}
\mathbf{G}=
\begin{pmatrix}
D_z/\epsilon \\ 0 
\end{pmatrix}\hat{j}
, \hspace*{0.3cm}
\label{eq:subsys}
\end{equation}
with $\mathbf{S}=0$ for simplicity, and $\hat{i}, \hat{j}$ being unit vectors along the $+x$ and $+y$ directions respectively. Conservation form for this subsystem came then be written in Cartesian co-ordinates as,
\begin{equation}
\frac{\partial\mathbf{U}}{\partial t}+\nabla\cdot\mathcal{F} = \frac{\partial\mathbf{U}}{\partial t}+\frac{\partial\mathbf{F}}{\partial x}+\frac{\partial\mathbf{G}}{\partial y}=0,
\label{eq:consforsub}
\end{equation}
where the flux is given as
\begin{equation*}
\mathcal{F}=[\mathbf{F},\mathbf{G}].
\end{equation*}
Writing the vector equation eq. \eqref{eq:consforsub} component-wise,
\begin{equation}
\frac{\partial}{\partial t}\begin{pmatrix}
B_x \\ B_y
\end{pmatrix}
+\frac{\partial}{\partial x}\begin{pmatrix}
0 \\ -D_z/\epsilon
\end{pmatrix}
+\frac{\partial}{\partial y}\begin{pmatrix}
D_z/\epsilon \\0
\end{pmatrix}=0.
\end{equation}
\begin{equation}
\frac{\partial}{\partial t}\begin{pmatrix}
B_x \\ B_y
\end{pmatrix}
+\begin{pmatrix}
\frac{\partial D_z/\epsilon}{\partial y} \\ \\
          -\frac{\partial D_z/\epsilon}{\partial x}
\end{pmatrix}=0.
\label{eq:continuousvector}
\end{equation}
A representation of eq. \eqref{eq:continuousvector} as a time-dependent PDE, is given by
\begin{equation}
\frac{\partial \mathbf{U}}{\partial t} = \mathbf{R}\left(\mathbf{U}\right),
\label{eq:abstractcontinuous}
\end{equation}
where $\mathbf{R}$ is the spatial operator. Consider a higher order discretization of eq. \eqref{eq:consforsub} of the form \cite{KompenhansRubio},
\begin{equation}
\int_\Omega \nabla\cdot\mathcal{F}\left(\mathbf{U}\right)\psi d\mathbf{x}=0 \rightarrow \sum_{\mathbb{D}^k\in \Omega_h}\mathbf{R}_p\left(\mathbf{U}_p\right)=0,
\end{equation}
where $\psi$ is a test function and $p$ is the degree of the polynomial bases used to represent the discrete solution $\mathbf
{U}_p$. $\mathbf{R}_p$ represents the discrete spatial partial differential operator. A lower $q$th order approximation on the same mesh can be formed by projecting the $p$-th order accurate solution to a space of $q$-th order basis functions,
\begin{equation}
\tilde{\mathbf{U}}_p^q=I_p^q\mathbf{U}_p,
\end{equation}
where $I_p^q$ is the transfer operator with $p>q$, and $\left(\tilde{\cdot}\right)$ is a restricted higher order (here, $p$-th order) quantity.
\par A relative truncation error \cite{Rueda-Ramirez2019,Fraysse2012} between levels $p$ and $q$,  is usually defined as,
\begin{equation}
\bm{\tau}_p^q=I_p^q\mathbf{R}_p\left(\mathbf{U}_p\right)-\mathbf{R}_q\left(I_p^q\mathbf{U}_p\right)= \llbracket \mathbf{R}\left(\mathbf{U}\right)\rrbracket_p^q.
\label{eq:definition_tau}
\end{equation}
Here, the notation $\llbracket \left(\cdot\right) \rrbracket_p^q$ represents the difference between quantities at refinement levels $p$ and $q$ as shown. Operator $\mathbf{R}$ of varying spatial accuracies acts on pure and restricted solution vectors $\mathbf{U}$, via the transfer operator $I_p^q$ restricting the higher ($p$-th) order solution to a lower ($q$-th) order. This relative truncation error is commonly used in multigrid and multilevel techniques for explicit time marching schemes \cite{ChatCiCP}, as a forcing function for maintaining higher ($p$-th) order solution while operating at a lower ($q$-th) order. Similarly, a relative divergence error \cite{ApurvaDiv} is defined as
\begin{equation}
\gamma_p^q=\bm{\nabla}\cdot\tilde{\mathbf{B}}_p^q-\bm{\nabla}\cdot \mathbf{B}_{q}
\label{eq:def_gamma}
\end{equation}  The relative truncation and relative divergence errors are related to each other via the Divergence Error Evolution Equation (DEEE), which in continuous form is derived in \cite{ApurvaDiv} to be,
\begin{equation}
\boxed{\frac{d\gamma_p^q}{dt}=\bm{\nabla}\cdot\bm{\tau}_p^q}.
\label{eq:causal}
\end{equation}
Eq. \eqref{eq:causal} expresses compactly that the relative truncation error acts as a source for the relative divergence error.   For $p\rightarrow\infty$, \textit{i.e.} corresponding to the continuous case, then eq. \eqref{eq:causal} suggests that it is due to the act of discretization that a divergence error is generated.  
The divergence error is contained in the discrete solution, and is not a result of the finite accuracy of the divergence operators used. This is emphasized above by defining $\gamma_p^q$ using infinitely accurate divergence operators. 
\par Consider a plane wave solution to eq. \eqref{eq:TM}, restricted to a single frequency  $\omega$ here, as a representative instance \cite{Hesthaven},
\begin{subequations}
\label{eq:planewavesoln}
\begin{gather}
D_z= D_{z0}e^{(\mathbf{k}\cdot\bm{x}-\omega t)}, \\
B_y=-\sqrt{\frac{\mu}{\epsilon}}D_z cos\phi, \\
B_x=\sqrt{\frac{\mu}{\epsilon}}D_z sin\phi,
\end{gather}
\end{subequations}
where $\bm{x}$ is a position vector in the $xy$ plane; $\mathbf{k}=\frac{\omega}{c}\left[\cos\phi, \sin\phi \right]^T$ is the wavenumber with $\phi$ being the angle made with the $+x$ axis.  $c=\frac{1}{\sqrt{\mu\epsilon}}$ is the speed of propagation in the medium. 
In this case, $\bm{\tau}_p^q$ defined as eq. \eqref{eq:definition_tau} evaluates to,
\begin{equation}
\bm{\tau}_p^q= \llbracket\mathbf{R}\left(\mathbf{U}\right) \rrbracket_p^q= \begin{pmatrix}
		   \left\llbracket\frac{\partial {B_y}}{\partial y}\right\rrbracket_p^q \frac{c}{cos\phi} \\ \\
   		   \left\llbracket\frac{\partial B_x}{\partial x}\right\rrbracket_p^q \frac{c}{sin\phi}
   		   \end{pmatrix}
= \begin{pmatrix}
		   -\left\llbracket\frac{\partial {E_z}}{\partial y}\right\rrbracket_p^q \\ \\
   		   \left\llbracket\frac{\partial E_z}{\partial x}\right\rrbracket_p^q
   		   \end{pmatrix}.
\label{eq:taupq}
\end{equation}
A relevant simplification for such a monochromatic plane wave solution using a fully discrete relative divergence error  $\hat{\gamma}_p^q$ is given in \cite{ApurvaDiv} by,
\begin{equation}
\boxed{\hat{\gamma}_p^q =\bm{\tau}_p^q\cdot\frac{\bm{\phi}}{c}},
\label{eq:scalarproxy}
\end{equation}
where $\bm{\phi}=\left[\cos\phi, \sin\phi \right]^T$, is a unit vector along the direction of wave travel. The fully discrete relative divergence error $\hat{\gamma}_p^q$ is defined as,
\begin{equation}
\hat{\gamma}_p^q =I_p^q\bm{\nabla}_p\cdot\mathbf{B}_p-\bm{\nabla}_q\cdot\left(I_p^q\mathbf{B}_p\right)
=\llbracket \bm{\nabla}\cdot\mathbf{B} \rrbracket_p^q,
\label{eq:definition_gammahat}
\end{equation}
which uses discrete divergence operators and includes the errors associated with them. The definition eq. \eqref{eq:definition_gammahat} is a practically computable relative divergence error defined as eq. \eqref{eq:definition_tau} \cite{ApurvaDiv}. $\hat{\gamma}_p^q$ is given component-wise by,
\begin{equation}
\hat{\gamma}_p^q = \llbracket \bm{\nabla}\cdot\mathbf{B} \rrbracket_p^q
= \left\llbracket\frac{\partial B_x}{\partial x}\right\rrbracket_p^q
+ \left\llbracket\frac{\partial {B_y}}{\partial y}\right\rrbracket_p^q.
\label{eq:gammapq}
\end{equation}
Relation \eqref{eq:scalarproxy}  forms the essential link between the two errors and institutes $\hat{\gamma}$ as a scalar proxy of $\bm{\tau}$.
Since the divergence error of the solution is readily available, it serves as an inexpensive driver in adaptive algorithms having the typical ease of computation of feature-based methods. However, these mutual relations make this indicator, also a rigourous truncation error-based indicator. The truncation and discretization errors are mutually related through the Discretization Error Transport Equation (DETE), which shows that truncation error acts as a local source of discretization error \cite{KompenhansRubio}. This made the case for preferring truncation error as a sensor in a number of adaptive algorithms for hyperbolic problems in literature. Hence, the extensive work in the area of truncation error estimation \cite{Rueda-Ramirez2019,KompenhansRubio,Syrakos2012}. 

\subsection{Comparison between local gradient and divergence errors}
 In section \ref{sec:Convergence_EME}, we established EM energy as a feature-based adaptivity driver, heuristically adapting based on $\nabla E$ (subscript $z$ dropped, considering the $TM_z$ mode). Here, we use error in gradient of local field to potentially develop a non-heuristic, error-driven indicator, similar to the use of error in SED in solid mechanics, shown in eq. \eqref{eq:SEDreln}. 
For comparison with $\gamma_p^q$, consider a relative gradient error in local field $E$, 
\begin{equation}
\bm{\nabla} \tilde{E}_p^q-\bm{\nabla}E_q=
\llbracket \bm{\nabla}E \rrbracket_p^q = 
\begin{pmatrix}
\left\llbracket \frac{\partial E}{\partial x} \right\rrbracket_p^q \\ \\ \left\llbracket \frac{\partial E}{\partial y} \right\rrbracket_p^q
\end{pmatrix}.
\label{eq:graderror}
\end{equation} 
In this context, both gradient and divergence of the solution are smoothness indicators, made of a combination of first spatial derivatives of the solution, as shown in eqns. \eqref{eq:gammapq} and \eqref{eq:graderror}. Despite this similarity, derivatives taken in the combination of divergence has an advantage over a simple gradient in the present context. From eqns. \eqref{eq:taupq} and \eqref{eq:graderror}, it can be seen that despite the magnitudes of the relative truncation and relative gradient errors being equal, a causal relation like the one with the relative divergence error in eq. \eqref{eq:scalarproxy} does not follow. 
\par More importantly, to compute the relative gradient error between two discretization levels, no reference value for $\nabla \tilde{E}_p^q$ is readily available limiting the use of gradient of the local field to a heuristic indicator as shown in sec. \ref{sec:Convergence_EME}. In solid mechanics, this problem of finding a reference value for SED $\Psi^*$ (refer eq. \eqref{eq:SEDreln}) is overcome by estimating it, using the SED $\tilde{\Psi}$, obtained from a discrete solution followed with projection techniques. The practice of adopting an enhanced or smoothed field as a true field  on the basis of the obtained numerical solution, is well established in solid mechanics \cite{Hernandez1997,Tessler,HintonCampbell,Blacker}.
Unlike in solid mechanics, the solenoidal constraint in the time-domain Maxwell's equations means that a known datum (of zero divergence) is available at all points in space and time. Eq. \eqref{eq:causal} establishes relative divergence error as a proxy to the relative truncation error and, the condition of null divergence makes the divergence corresponding to any single discretization level, a relative divergence error. 
\par Therefore, unlike the relative gradient error, a divergence error based sensor need not rely on expensive estimation procedures and can be accurately computed. Moreover, it leads to a non-heuristic method eliminating the subjectivity involved with gradient-driven adaptation schemes. The divergence error-based indicator utilizes the structure of the governing equations eliminating the need to estimate reference states. It shows all desirable properties of adaptivity indicators identified in sec. \ref{sec:SEDprops}, along with the added advantage of a known reference. 
\par  As pointed out earlier in the introduction, the high cost of computation of an error indicator, and accommodating tedious computations in an existing code structure are limiting factors for the practical use of adaptive algorithms in commerical software \cite{ZienFunds}. Since computing divergence only uses derivatives of the solution vector, requisite routines are usually already available in an existing code, making it practically feasible to incorporate such an indicator. It is co-ordinate independent and a scalar, and combines the rigour of truncation error driven methods, with the practicality of feature-driven methods.
\par  The effort to advance from heuristic feature-driven methods to fully automated, rigorous error-driven adaptive methods seen in solid mechanics, is made here for the time-domain Maxwell's equations. We construct an adaptation routine based on the gradient of EM energy, followed by another, utilizing a divergence error indicator. 
\par The arguments presented thus far, are summarized as follows:
\begin{enumerate}
\item Like gradient of SED in solid mechanics, gradient of EM energy has been used for the time-domain Maxwell's equations, both leading to heuristic, feature-based adaptive schemes.
\item To advance from a heuristic feature driven method to an automated error-driven adaptive method, error in SED is used as an estimate of the discretization error in solid mechanics. With the hyperbolic time-domain Maxwell's equations, use of gradient of EM energy is limited to a feature-based method since it neither shows a direct relation with the relative truncation error, nor is economically computable.
\item A re-interpretation of the divergence error, inherent to any (except FDTD and dedicated divergence-free methods) numerical solution of the time-domain Maxwell's equations, establishes it as a proxy to the relative truncation error.
\item Along with having all desirable properties identified from the well-established SED-based indicators in solid mechanics, the divergence error has an additional advantage of being exactly computable, eliminating typical estimation procedures.
\end{enumerate}

\section{Results}
\label{sec:Results}
We present results for 3 testcases in this section - scattering from a PEC circular cylinder, a semi-open cavity, and a case of two adjacent PEC circular cylinders. These problems represent complexities arising from scatterers with various geometries and configurations. Uniform $p$ solutions are compared to the adaptive $p$ solutions for accuracy, along with a comparison of computational performance between the two adaptivity drivers: one based on gradient of the local field, and the other on divergence error of the solution. For a fair comparison, both adaptive methods use the same algorithm with respective local indicators as input to map them to a $p$-distribution. The algorithm takes a logarithm of the local data, segregating it into different orders of magnitude, and maps the highest value to a preset maximum allowable $p$, which is taken to be $8$, and the lowest to $1$ with all intermediate values mapped linearly. Hence, every node gets mapped to a proposed $p$ for the next iteration and the maximum $p$ amongst all nodes within an element is chosen. The elements forming the PML layer to truncate the domain boundaries, are manitained at the highest, \textit{i.e.} $8$-th order to resolve the rapidly decaying fields. For a thorough description, we refer the reader to \cite{ApurvaDiv}.
\par In EM wave scattering problems, a scattered wave is formed at the PEC boundary and propagated away from it. Thus, the scatterer acts as a local source resulting in a relatively higher divergence error close to its boundaries. Also, as the wave progresses in space, similar behaviour is expected along the foremost, or leading wavefront. Hence, we expect the divergence-driven adaptive method to allocate denser close to the scattering surface and the leading wavefront. Likewise, the higher relative gradients close to the scattering body due to its being the source, is expected to cause the gradient-based adaptive method to allocate denser. Additionally, unlike divergence, the harmonic scattered wave advecting in space, is expected to have higher relative gradients in the region between the scatterer and the leading wavefront as well, especially in the shadow region, as is typical in scattering problems. Hence, the distribution of $p$ across elements obtained using the gradient-driven adaptive method, is expected to qualitatively follow the contour of the local field, refining (by raising local $p$) along the boundary of the scatterer and extending it upto the leading wavefront.

\subsection{Scattering off a circular cylinder} 
\label{problem:cylscat}

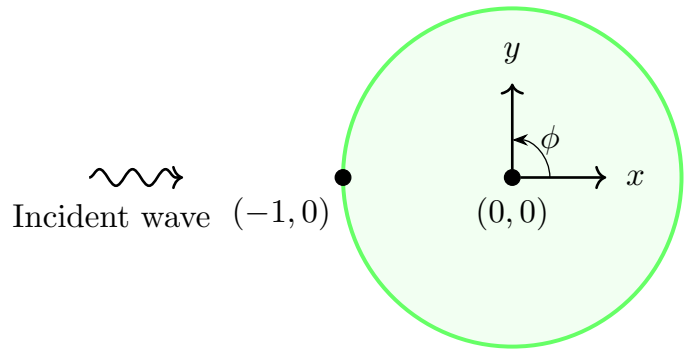
\begin{figure}
\centering
\scalebox{1.25}{
\begin{tikzpicture}[
photon/.style={decorate, decoration={snake}, draw=black},
squaregreennode/.style={circle, draw=black, fill=black, very thick, scale=0.5},
roundnode/.style={circle, draw=green!60, fill=green!5, very thick, minimum size=36mm},
roundnode2/.style={draw,shape=circle,fill=black,scale=0.5},
coord/.style={draw,shape=circle,fill=black,scale=0}
]

\node[roundnode]      (scat)         {};
\node[roundnode2,
label={[label distance=1.0cm]0:$x$},
label={[label distance=1.0cm]90:$y$},
label={[label distance=0.0cm]270:$\left(0,0\right)$},
label={[label distance=3cm]183:Incident wave},draw]      (orig) {};
\node[roundnode2,label={267:$\left(-1,0\right)$}] (leftorig) at (-1.8,0) {};
\node[] (phi) at (0.4,0.4) {$\phi$};

\draw[black, thick,->] (orig) -- (1,0);
\draw[thick,->,photon] (-4.5,0) -- (-3.5,0) ;
\draw[black, thick,->] (orig) -- (0,1);
\draw[thin, -Stealth] (.4,0) arc  (0:90:0.4);

\end{tikzpicture}}\\
\caption{Schematic for the circular cylinder scattering problem}
\label{fig:schematic}
\end{figure}

The two proposed drivers are applied to the problem of scattering from a PEC cylinder for validation. A schematic of the problem is shown in fig. \ref{fig:schematic}. Variations of the problem in terms of electrical sizes and incident TM and TE illumination are presented.  

\begin{figure}
\centering
\includegraphics[width=0.8\textwidth]{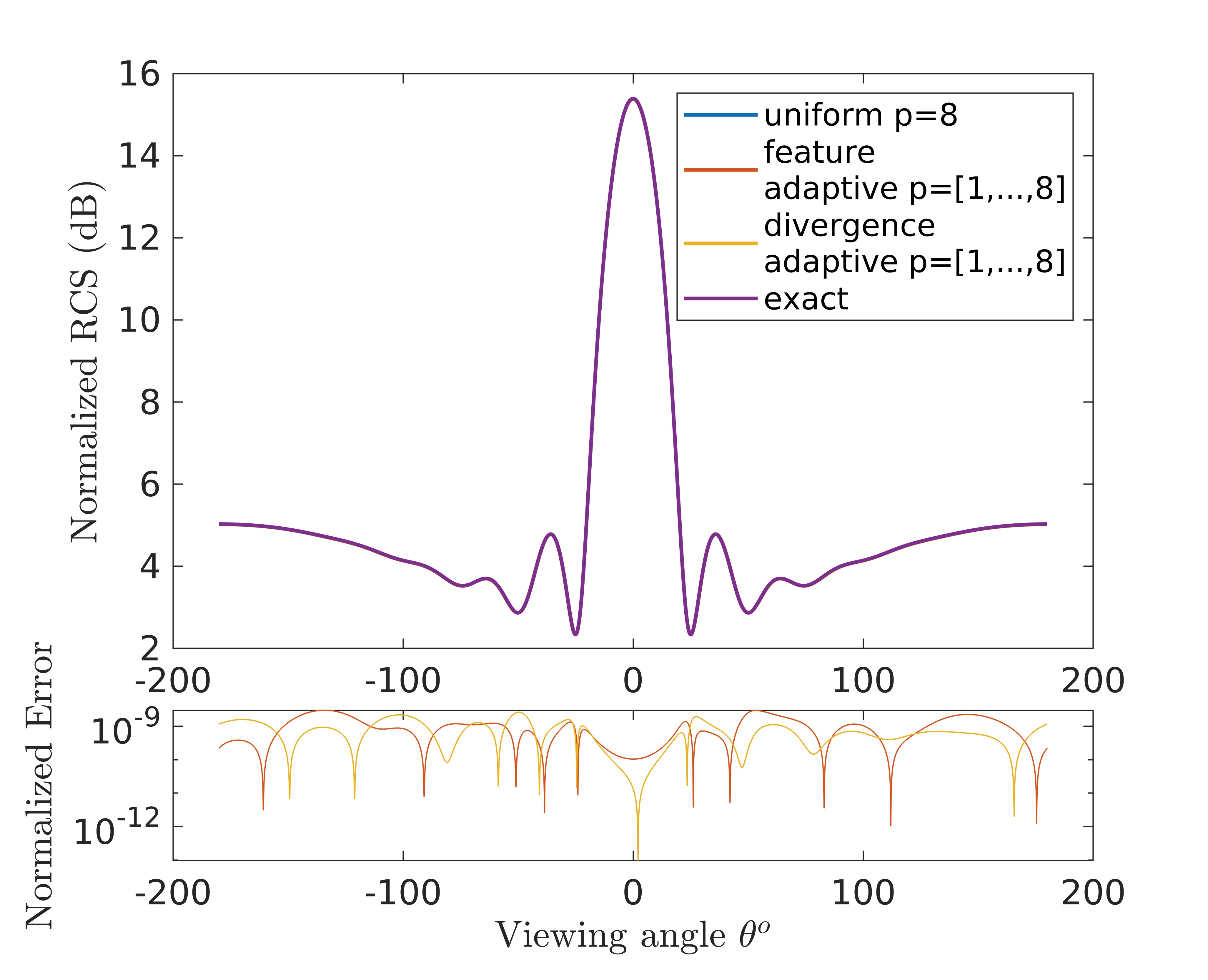}
\caption{RCS, circular cylinder scattering,  size $2\lambda$, TM illumination}
\label{fig:rcs2TM}
\end{figure}

\begin{figure}
\centering
\includegraphics[width=0.8\textwidth]{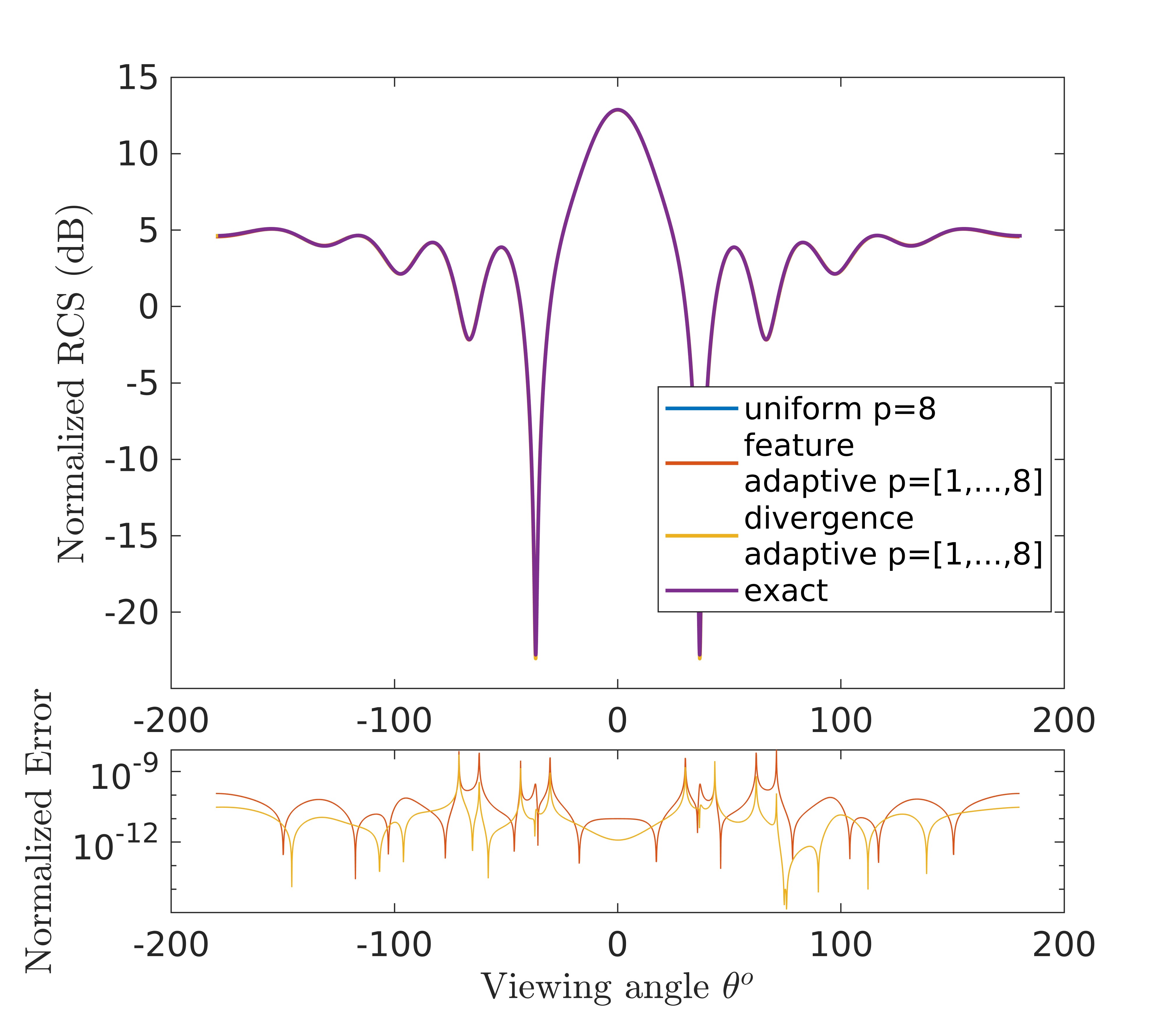}
\caption{RCS, circular cylinder scattering,  size $2\lambda$, TE illumination}
\label{fig:rcs2TE}
\end{figure}

\begin{figure}
\centering
\includegraphics[width=0.8\textwidth]{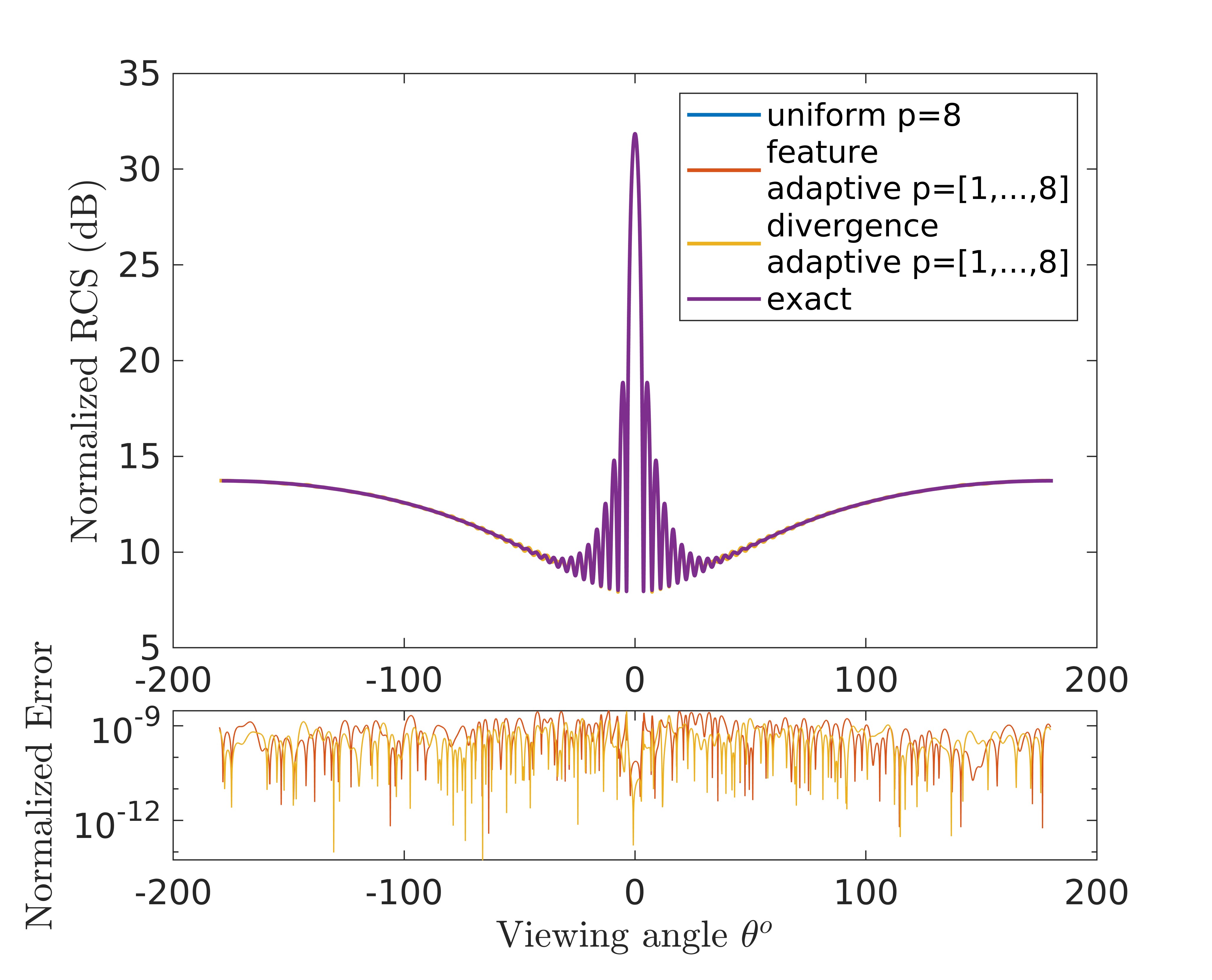}
\caption{RCS, circular cylinder scattering,  size $15\lambda$, TM illumination}
\label{fig:rcs15TM}
\end{figure}

The scattering width obtained numerically, compared to the exact solutions are shown in figs.\ \ref{fig:rcs2TM} and \ref{fig:rcs2TE}, under TM and TE illuminations respectively, for a scatterer $2\lambda$ in diameter. Fig. \ref{fig:rcs15TM} shows another instance with a larger scatterer, $15\lambda$ in size. 4 lines plots overlay each other including a uniform order method to compare against the 2 adaptive methods. For this canonical case, the exact solution is known and is plotted for comparison. Since the plots are visually indistinguishable, the difference between the RCS obtained using the adaptive $p=[1,\cdots,N]$ and uniform $p=N$ solutions is also shown. Viewing angle $\theta$ referred to in these plots, is measured from the $+x$ axis as depicted in the schematic fig.\ \ref{fig:schematic}. The adaptation routine operates after every iteration throughout the duration of the problem, including an initial unsteady phase, evolving into a harmonically steady state.

\begin{sidewaysfigure}
\flushleft
\includegraphics[width=\textwidth]{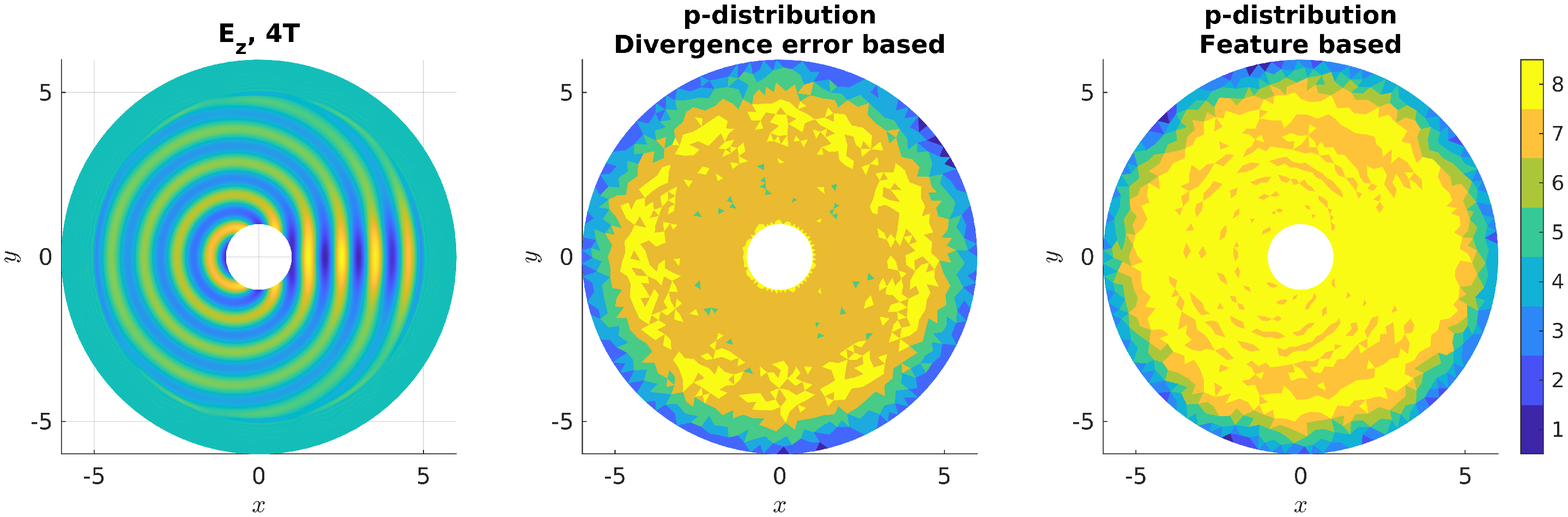}
\caption{Comparison between $p$ distributions, circular cylinder scattering, size $2\lambda$, TM illumination}
\label{fig:pdis1cyl}
\end{sidewaysfigure}

To demonstrate the dynamic behaviour of the two adaptation criteria, fig.\ \ref{fig:pdis1cyl} shows the scattered $E_z$ at the 4 time-period mark and the corresponding distributions of $p$ resulting from the adaptation. The divergence-based identifies local sources of divergence at the scattering surface, and along the leading wavefront travelling away from the body. This becomes significant in case of mutually interacting waves from multiple scattering surfaces, presented in later sections. Unlike divergence, the gradients do not decay away from the scatterer resulting in the feature-based method allocating higher order elements more densely as compared to that of the divergence error based driver since the gradients do not localize based strictly on the truncation error. The $p$-distribution pattern follows the contour of $E_z$ instead.

\subsection{Semi-open cavity}
\label{problem:SOcavity}
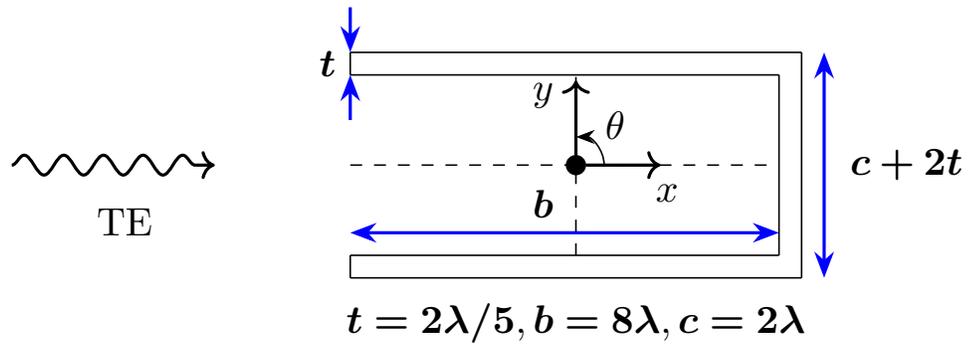
\begin{figure}
\centering
\scalebox{1.5}{
\begin{tikzpicture}[
photon/.style={decorate, decoration={snake}, draw=black},
roundnode2/.style={draw,shape=circle,fill=black,scale=0.5}]
\draw[] (-2,-1) -- (2,-1);
\draw[thick, {Stealth}-{Stealth},blue] (-2,-0.6) --node [pos=0.45,above,black] {$\bm{b}$} (1.8,-0.6);
\draw[] (-2,1) -- (2,1);
\draw[] (-2,-0.8) -- (1.8,-0.8);
\draw[] (-2,0.8) -- (1.8,0.8);
\draw[] (-2,-1) -- (-2,-0.8);
\draw[] (-2,0.8) -- (-2,1);
\draw[] (2,-1) -- (2,1);
\draw[thick, {Stealth}-{Stealth},blue] (2.2,-1) --node [midway,right=0.1cm,black] {$\bm{c+2t}$} (2.2,1);
\draw[] (1.8,-0.8) -- (1.8,0.8);
\draw[thick, -{Stealth},blue] (-2,1.4) -- (-2,1);
\draw[thick, -{Stealth},blue] (-2,0.4) -- (-2,0.8);
\node[] (t) at (-2.2,0.9) {$\bm{t}$};
\draw[thin, dashed] (-2,0) -- (1.8,0);
\draw[thin, dashed] (0,-0.8) -- (0,0.8);
\draw[thick,->,photon] (-5,0) -- (-3.2,0);
\draw[thin, -Stealth] (0.25,0) arc  (0:90:0.25);
\node[] (theta) at (0.35,0.35) {$\theta$};
\node[] (measurements) at (0,-1.4) {$\bm{t=2\lambda/5}, \bm{b=8\lambda}, \bm{c=2\lambda}$};
\draw[black, thick,->] (orig) -- (0.75,0);
\draw[black, thick,->] (orig) -- (0,0.75);
\node[roundnode2,
label={[label distance=0.5cm]-5:$x$},
label={[label distance=0.3cm]101:$y$},draw]      (orig) {};
\node[] (TE) at (-4.0,-0.5) {TE};
\end{tikzpicture}}
\caption{Schematic for the semi-open cavity problem}
\label{fig:schematic_SOcavity}
\end{figure}

\begin{figure}
\centering
\includegraphics[width=1.2\textwidth]{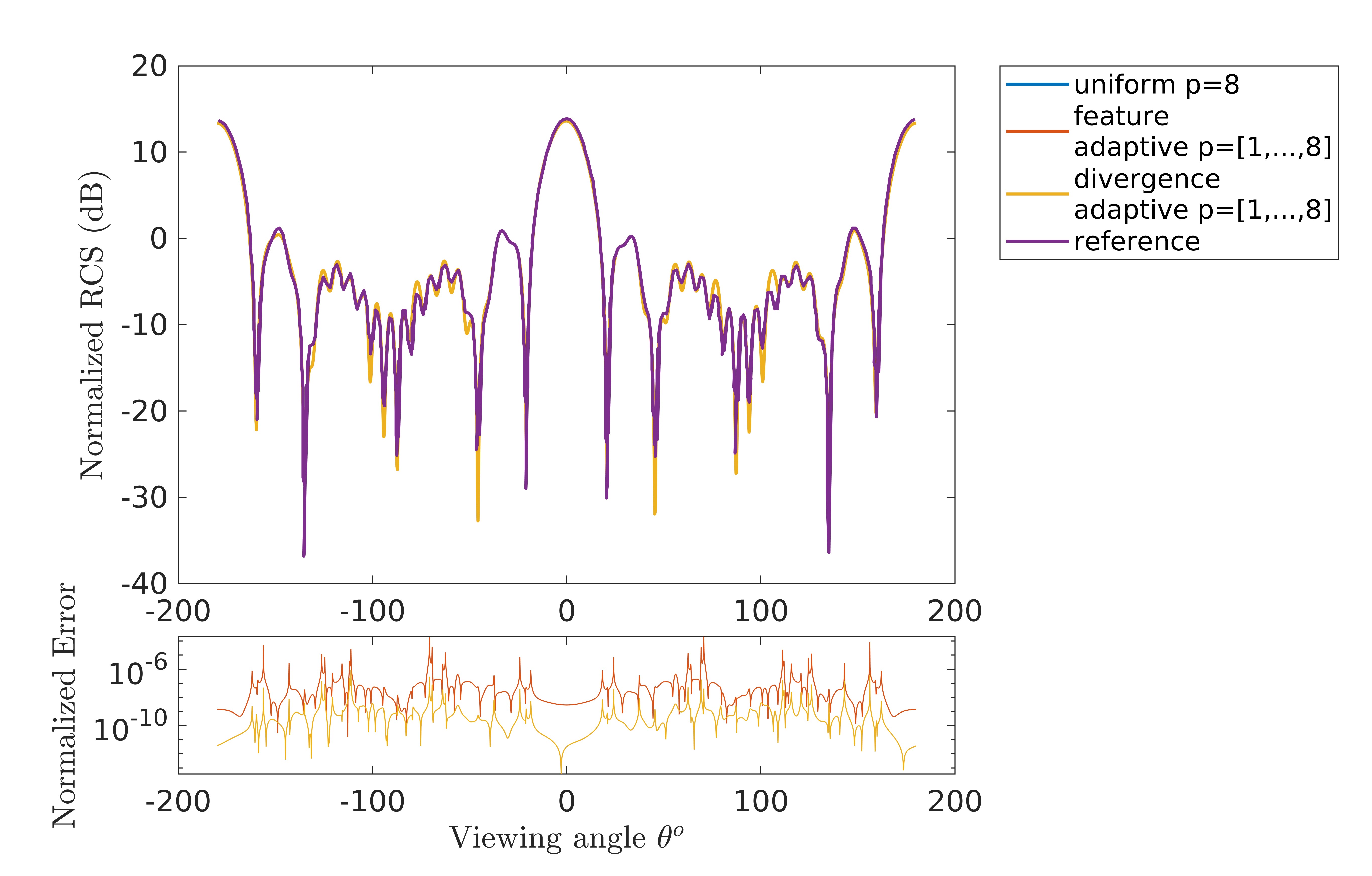}
\caption{RCS, semi-open cavity problem.}
\label{fig:rcsSOcavity_phi0}
\end{figure}

Fig.\ \ref{fig:schematic_SOcavity} shows a schematic of the semi-open cavity problem. A TE monochromatic plane wave travels towards the closed-end of the cavity. The computational domain includes a PML 1$\lambda$ wide at the outer boundaries to truncate the domain.  \ref{fig:rcsSOcavity_phi0}. The overlaid RCS plots in fig. \ref{fig:rcsSOcavity_phi0} show that the adaptive solutions are essentially as accurate as the uniform $p$ solution. A $5$th order accurate hybrid Galerkin solution to this problem, presented in \cite{Davies2009} has been used as reference. Results shown have been computed at the 40 period mark, the solution being well converged.

\begin{sidewaysfigure}
\flushleft
\includegraphics[width=1.1\textwidth]{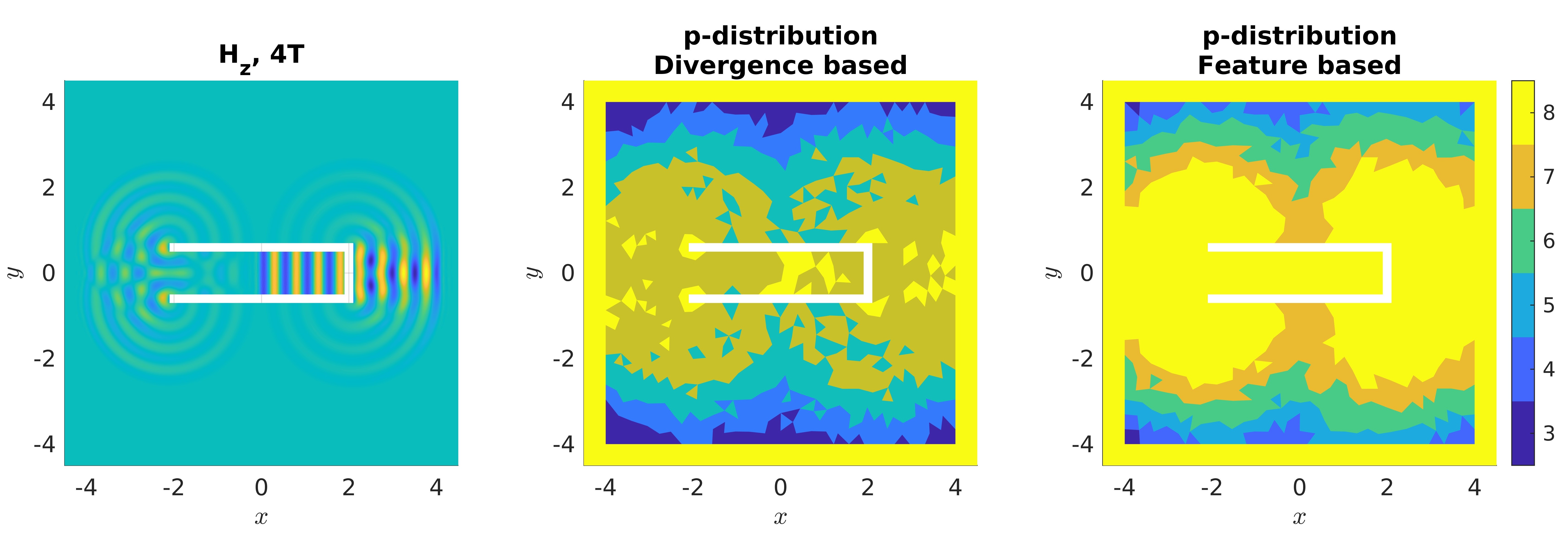}
\caption{Comparison between $p$ distributions, semi-open cavity problem}
\label{fig:pdisSOcavity}
\end{sidewaysfigure}

We compare the $p$-distributions at an intermediate time level of 4 periods in the unsteady regime to highlight key differences between the two adaptation schemes, in fig. \ref{fig:pdisSOcavity}. The scattered wave in the cavity has reached halfway length and this raises the local divergence near this region, as also at the corners, seen in the divergence-based $p$-distribution. The contours of $H_z$ also show that the scattered waves emerging at both open and closed ends of the cavity (along $x$-axis) completely surround the scattering surface. This creates higher relative gradients around the scattering surface resulting in denser allocation by the feature-based method, compared to the  divergence error based, that localizes most higher order elements along the scattering surfaces.

\subsection{Two adjacent cylinders}
\label{sec:2cyl}
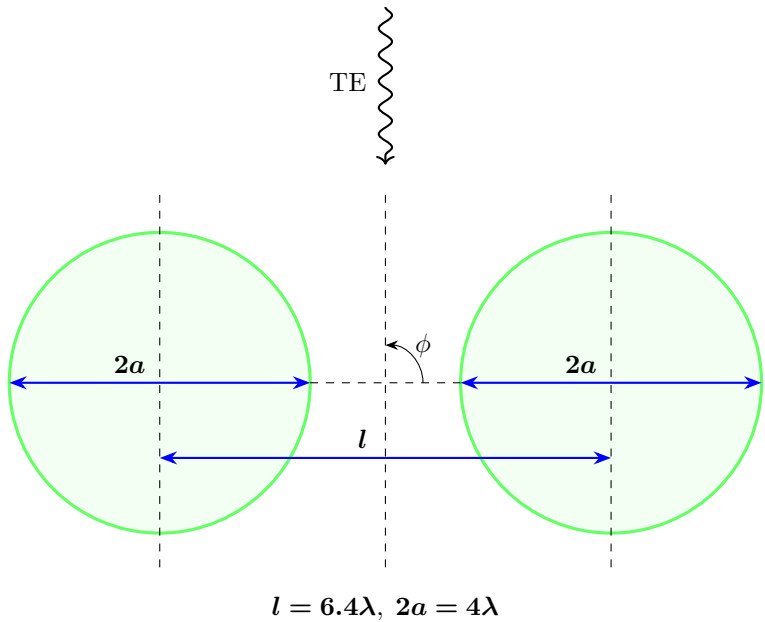
\begin{figure}
\centering
\scalebox{1}{
\begin{tikzpicture}[
photon/.style={decorate, decoration={snake}, draw=black}]
\node[] (orig) at (-0.5,4) {TE};
\node[] (phi) at (0.5,0.5) {$\phi$};
\node[] (l) at (0,-3) {$\bm{l=6.4\lambda},\hspace{0.1cm}\bm{2a=4\lambda}$};
\filldraw[color=green!60, fill=green!5, very thick](-3,0) circle (2);
\filldraw[color=green!60, fill=green!5, very thick](3,0) circle (2);
\draw[thick,->,photon] (0,5) -- (0,2.9);
\draw[thin, dashed] (-3,2.5) -- (-3,-2.5);
\draw[thin, dashed] (3,2.5) -- (3,-2.5);
\draw[thin, dashed] (0,2.5) -- (0,-2.5);
\draw[thin, dashed] (-1,0) -- (1,0);
\draw[thin, -Stealth] (.5,0) arc  (0:90:0.5);
\draw[thick, {Stealth[scale=1]}-{Stealth[scale=1]},blue] (-3,-1) --node [pos=0.45,above,black] {$\bm{l}$} (3,-1) ;
\draw[thick, {Stealth[scale=1]}-{Stealth[scale=1]},blue] (-5,0) --node [pos=0.4,above,black] {$\bm{2a}$} (-1,0);
\draw[thick, {Stealth[scale=1]}-{Stealth[scale=1]},blue] (1,0) --node [pos=0.4,above,black] {$\bm{2a}$} (5,0);
\end{tikzpicture}}
\caption{Schematic for the scattering off 2 adjacent cylinders problem}
\label{fig:schematic_2cyl}
\end{figure}

Another canonical problem that addresses the complexity of multiple reflections and reciprocal interactions is that of multiple scatterers \cite{YoungBertrand}. Here, we investigate scattering from two adjacent circular cylindrical PEC scatterers, the schematic for which is shown in fig. \ref{fig:schematic_2cyl}. An incident TE wave illuminates the two cylinders travelling along the $-y$ axis. The distance separating the scatterers ($6.4$ wavelengths) and the diameter of each ($4$ wavelengths) makes it lie in the optical scattering range. The circular domain is truncated with a PML $1\lambda$ wide with the interface located at a radius of 7.5, and the outermost boundary at radius 8.5. The solution takes 50 periods to converge and a reference RCS plot is taken from \cite{Davies2009} that used a $6$th order accurate hybrid Galerkin solution.

\begin{figure}
\centering
\includegraphics[width=0.8\textwidth]{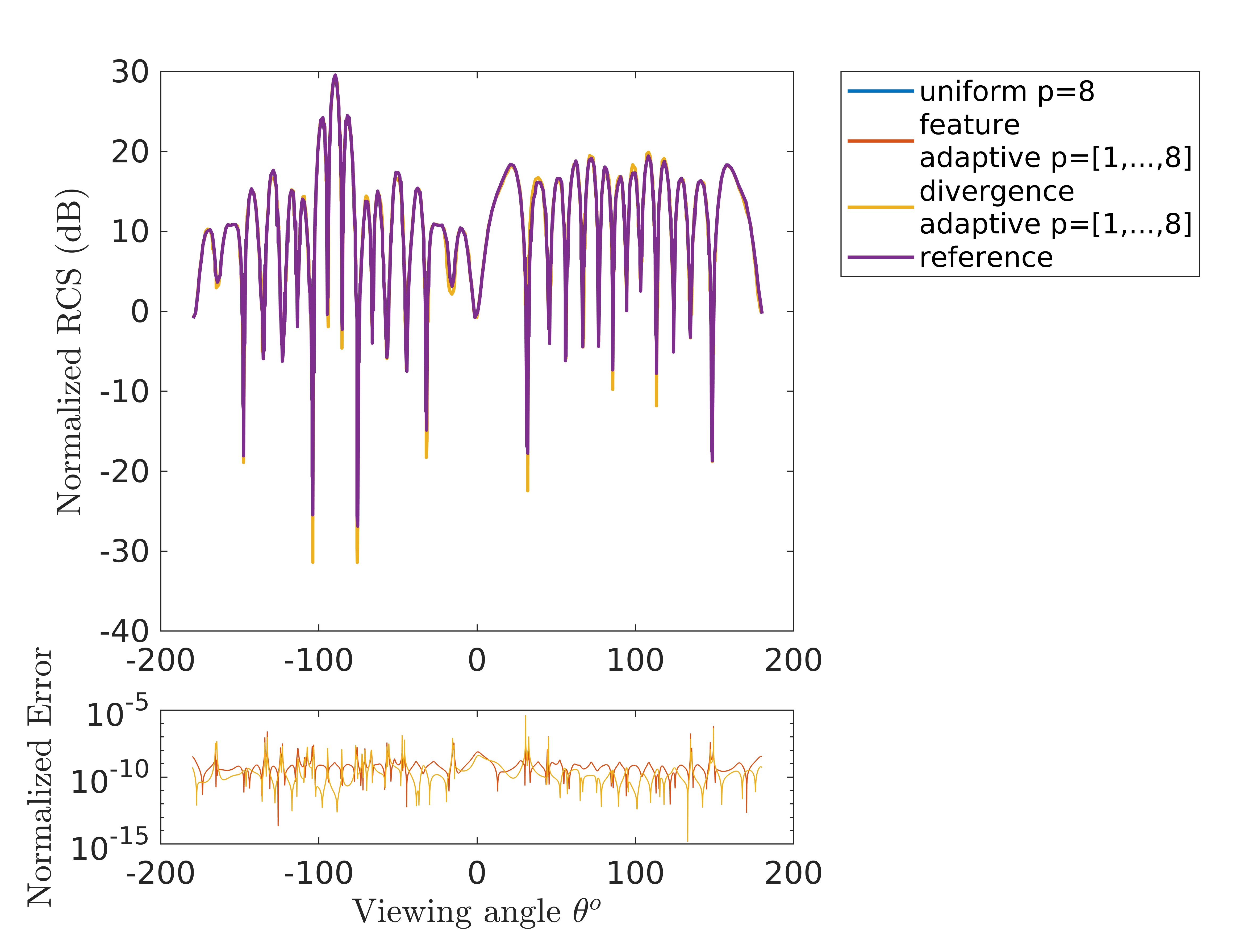}
\caption{RCS, 2 adjacent cylinders problem}
\label{fig:rcs2cyl}
\end{figure}

Continuing with the pattern of presenting RCS results as in earlier sections, fig.\ \ref{fig:rcs2cyl} shows the uniform, adaptive and reference plots. The uniform and the adaptive solution plots are close to each other and in agreement with the reference.

\begin{sidewaysfigure}
\centering
\includegraphics[width=1.1\textwidth]{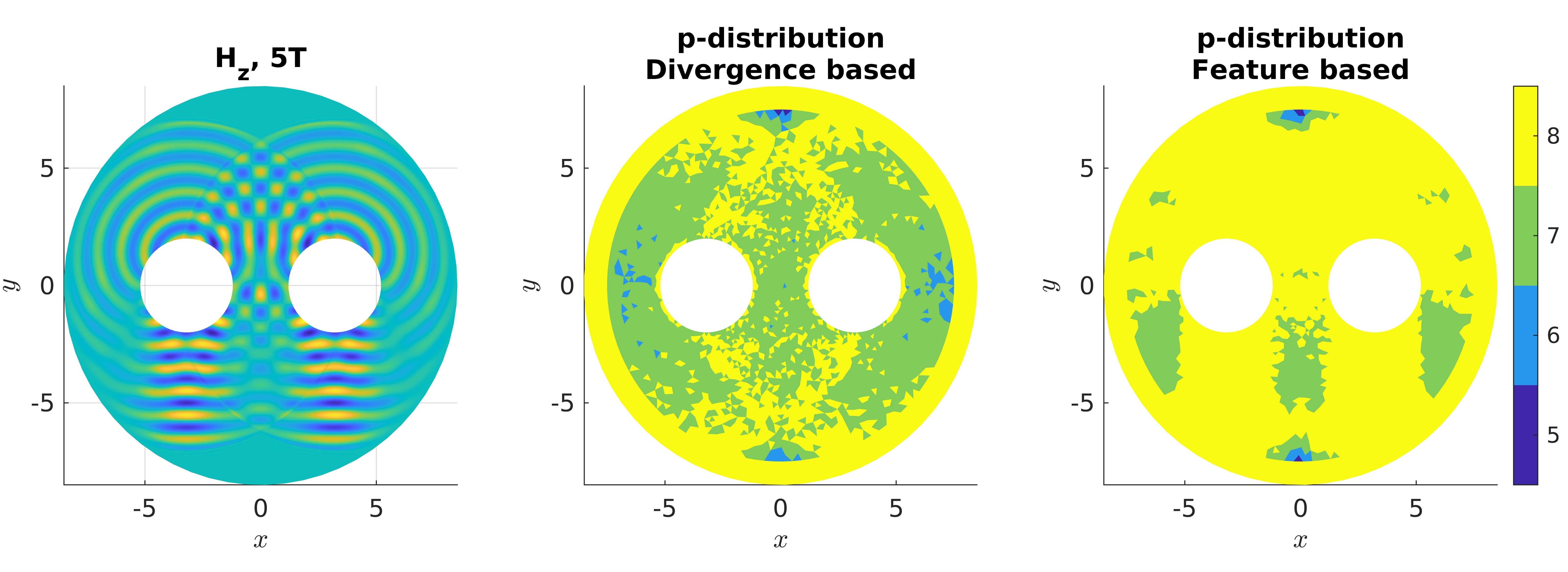}
\caption{Comparison between $p$ distributions, 2 adjacent cylinders problem}
\label{fig:pdis2cyl}
\end{sidewaysfigure}

Fig.\ \ref{fig:pdis2cyl} compares $p$-distributions at the 5 period mark, developed enough to show all key features. The contour on the lit side (the upper half) shows two leading wavefronts scattered off each cylinder, and also off the neighbouring cylinder. The divergence error driver captures these wavefronts in both the illuminated and shadow regions. The gradient-driven method on the other hand, follows the contours of the scattered field closely, along with a characteristic denser allocation as seen in previous cases.
\par For similar accuracies (determined by the resulting RCS from both methods), it is observed that the divergence error based method leads to a more economical allocation of $p$, concentrating the DOFs to local sources of divergence error, and in turn truncation errors. 
The scattering surface acting as a local source of divergence error, gets assigned a denser allocation of higher order elements relative to the region between the surface and the leading wavefront. Also, the divergence-error based method tracks the leading wavefront of the scattered wave from a body on its way to a neighbouring scatterer, thereby maintaining higher order accuracy in cases of mutually interacting waves between multiple scatterers.  On the other hand, a feature-based method driven by gradient of the local field, allocates denser not only close to the scattering surface, but also in the region upto the leading wavefront following the harmonic scattered wave. Hence, unlike divergence error, it results in over-compensating by concentrating DOFs closer to, but also away from local truncation error sources. This leads to a higher demand in DOFs, quantified in the following section.

\subsection{Computational performance}
\begin{table}
\begin{center}
\begin{tabular}{|c|c|c|c|c|}
\hline
\multirow{2}*{\textbf{Problem}}
& \multicolumn{2}{c|}{\textbf{Savings in DOFs (\%)}} 
& \multicolumn{2}{c|}{\textbf{effective $p$}} \\
\cline{2-5}
 & Divergence error  & Gradient of energy  & Divergence error  & Gradient of energy \\ \hline
1 cylinder, 2$\lambda$, TM & 43.14 & 31.36 & 5.31 & 6.07\\ 
1 cylinder, 2$\lambda$, TE & 43.41 & 31.34 & 5.30 & 6.07\\
1 cylinder, 15$\lambda$, TM & 34.36 & 20.86 & 5.90 & 6.72\\
1 cylinder, 15$\lambda$, TE & 35.01 & 20.84 &  5.87 & 6.73\\ 
 Semi-open cavity & 25.95 & 18.12 & 6.31 & 6.82\\
 2 adjacent cylinders & 23.39 & 13.77 & 6.57 & 7.23 \\ \hline
\end{tabular}
\end{center}
\caption{Computational savings in DOFs and effective $p$ employed to achieve accuracy corresponding to a uniform 8th order method.}
\label{table:comper}
\end{table}

In this section, we compare the computational gain from the two proposed adaptive methods. For equally accurate solutions, the two schemes evolve different $p$-distributions and the impact on computational savings is compared in table \ref{table:comper}. The savings in DOFs are compared to a uniform $p$ solution. The effective $p$ shown for individual testcases, is the $p$, averaged over time and space. It serves as a measure of the computational savings made in obtaining $8$-th order accurate solutions with methods using lower order spatial operators, in an average sense. While considerable savings are made in all testcases, the recurring theme of the feature-based scheme using denser allocations than the divergence error-based, follows from the $p$-distribution plots shown in earlier sections, and is quantified in table \ref{table:comper}.
  
\section{Conclusion}
An extension of the principles on which established drivers for adaptive methods in solid mechanics are based upon, is made to CEM for non-FDTD solutions to the time-domain Maxwell's equations. A feature-based method guided by gradient of local EM energy, and another using the divergence error are presented. The proposed methods use a combination of spatial operators locally varying in order of accuracy, to achieve higher order accurate solutions, using fewer degrees of freedom. Numerical results for canonical testcases of scattering problems are presented demonstrating the effectiveness of the proposed schemes. Theoretically, it is shown that the role strain energy plays in adaptive methods for solid mechanics of acting as a proxy to numerical error, EM energy in the current context cannot. Unlike strain energy, EM energy does not contain any more information than the solution and despite showing some desirable properties, is not a viable substitute. The use of EM energy as an adaptivity driver remains limited to only a heuristic, feature-based method. As an alternative, the relation between truncation and divergence errors is used to develop a divergence error driven adaptive algorithm. The highlight of using the proposed divergence error driver is that it needs no expensive error estimation algorithms and is easily computed. This is enabled by the knowledge of a reference state of null divergence, afforded in the time-domain Maxwell's equations, a  distinguishing feature from other applications. Divergence error makes for a robust indicator to develop automated, error-driven adaptive methods. The divergence error indicator not only satisfies all requisite properties that established error indicators in solid mechanics possess, it is easily incorporated in existing code structures. More importantly, it is highlighted that numerical divergence error when not detrimental to the underlying physics, may be utilized to a computational gain. 

\printbibliography
\end{document}